\documentstyle[12pt]{article}
\title{Violation of unitarity by Hawking radiation \\
does not violate energy-momentum conservation}
\author{Hrvoje Nikoli\'c \\
Theoretical Physics Division, Rudjer Bo\v{s}kovi\'{c} Institute, \\
P.O.B. 180, HR-10002 Zagreb, Croatia \\
{\normalsize e-mail: hnikolic@irb.hr} \\
\makebox[1in]{} \\
}
\date{\today}
\begin{document}
\maketitle
\begin{abstract}
An argument by Banks, Susskind and Peskin (BSP),
according to which violation of unitarity would violate either locality or energy-momentum conservation,
is widely believed to be a strong argument against non-unitarity of Hawking radiation.
We find that the whole BSP argument rests on the crucial assumption
that the Hamiltonian is not highly degenerate, and point out that this assumption is 
not satisfied for systems with many degrees of freedom.
Using Lindblad equation, we show that high degeneracy of the Hamiltonian allows local non-unitary
evolution without violating energy-momentum conservation.  
Moreover, since energy-momentum is the source of gravity, we argue that 
energy-momentum is necessarily conserved  for a large class of non-unitary systems with gravity.
Finally, we explicitly calculate the Lindblad operators for non-unitary Hawking radiation
and show that they conserve energy-momentum.
\end{abstract}
\vspace*{0.5cm}

\section{Introduction}

According to the semi-classical theory, Hawking radiation from black holes \cite{hawk1}
is a non-unitary process \cite{hawk2}.   
Nevertheless, there are many different attempts to restore unitarity of Hawking radiation
(see e.g. \cite{gid,har,pres,pag,gid2,str,math,hoss,harlow,fabbri} for reviews)
and the belief that the true black-hole evolution should be unitary is highly prevailing. 
Yet, neither of the attempts to restore unitarity is without intrinsic problems, 
and many of the attempts contradict each other. 
In a confrontation with these problems, it seems wise not yet to completely dismiss
the possibility that Hawking radiation might really be a non-unitary process.

One of the most frequent arguments for unitarity is authoritative reference
to the work by Banks, Susskind and Peskin (BSP) \cite{BSP}, who argued that 
violation of unitarity would violate either locality or energy-momentum conservation.
Some possible loopholes in BSP arguments have been suggested in the literature
\cite{hawk4,srednicki,liu,gielen,unruh-wald,unruh,oppenheim},
but neither of these loopholes seems to be taken seriously in a wider community.  

In this paper we find and exploit a new loophole in the BSP argument.
In Sec.~\ref{SEC-BSP} we critically re-examine the arguments used in \cite{BSP} and find
that one crucial assumption in their argument is not satisfied. This is the assumption
that the Hamiltonian is not highly degenerate, which we 
show in Sec.~\ref{SEC-degen} to be a wrong assumption.
In Sec.~\ref{SEC-open} we explain in detail why high degeneracy of the Hamiltonian allows
non-unitary evolution to be compatible with both energy-momentum conservation
and locality. 
Moreover, in Sec.~\ref{SEC-open-grav} we argue that energy-momentum conservation is not merely allowed,
but even mandatory, for a large class of non-unitary systems with gravity.
Finally, we apply these general results to the case of Hawking radiation
in Sec.~\ref{SEC-hawk}, and make some more general remarks in Sec.~\ref{SEC-disc}. 

\section{A critical re-examination of BSP arguments}
\label{SEC-BSP}

In this section we re-examine some of the crucial arguments used in \cite{BSP}, with emphasis
on those arguments which we find problematic.

BSP find that
a non-unitary evolution of the density matrix $\rho$ can be described by a differential equation
of the form
\begin{equation}\label{eq1}
 \dot{\rho}=-i[H_0,\rho]-\frac{1}{2}\sum_{\alpha,\beta\neq 0}
h_{\alpha\beta} (Q^{\beta}Q^{\alpha}\rho+\rho Q^{\beta}Q^{\alpha} -2Q^{\alpha}\rho Q^{\beta}) ,
\end{equation}
which is Eq.~(9) in their paper \cite{BSP}.
Here $Q^{\alpha}$ are unspecified hermitian operators, $h_{\alpha\beta}$ are 
unspecified c-numbers, and $H_0$ is an operator which can explicitly be expressed in terms of 
$Q^{\alpha}$.  
(Actually, they make a typo by omitting the $h_{\alpha\beta}$ factor in their Eq.~(9),
but it does not influence their further analysis.)
Then, at the beginning of their Section 4, they make the following argument:
\begin{quote}
``The failure of energy conservation can be seen from the following observation:
What if the theory did possess some hermitian operator $H$ (not necessarily equal
to $H_0$) which was conserved by the dynamics? Then any $\rho$ which was a function only of 
$H$ could not change under the action of (9). However, this is possible only if (9)
contains only operators $Q$ which are simultaneously diagonalizable with $H$. 
{\it Unless $H$ has highly degenerate eigenvalues (a property which would exclude it
as a good candidate for the energy)}, this is a serious restriction on $h_{\alpha\beta}$,
especially if $Q_{\alpha}$ must be a local operator rather than a global charge.''
(our italics)    
\end{quote}
After that argument, in the rest of their paper they study in detail these restrictions on $h_{\alpha\beta}$
and explain how these restrictions violate either locality or energy-momentum conservation.

The italicized part of the quote above represents the core of our criticism of the BSP argument.
Indeed, the italicized part shows that they are aware of a possible loophole for their argument; 
the argument is invalid if $H$ has highly degenerate eigenvalues. Nevertheless, 
the quote shows that they dismiss
this loophole by saying that it ``would exclude it as a good candidate for the energy''.
Unfortunately, they do not explain {\em why} a Hamiltonian with highly degenerate eigenvalues
would not be a good candidate for the energy and we cannot find any reasonable justification 
for such a claim. Just the opposite, we claim that
a good candidate for the Hamiltonian with many degrees of freedom {\em must} have highly degenerate eigenvalues. 

The physical reasons for high degeneracy of the Hamiltonian are explained in detail 
in Sec.~\ref{SEC-degen}, but let us here present a simple mathematical argument.
In a system with $N$ degrees of freedom, the Hilbert space is usually spanned 
by a set of basis vectors of the form
\begin{equation}\label{eq2}
 |\xi_1,\xi_2,\ldots,\xi_N\rangle ,
\end{equation}
parameterized by $N$ real (continuous or discrete) parameters $\xi_1,\xi_2,\ldots,\xi_N$.
One of the parameters, say $\xi_1$, usually represents the energy $E$, so that all other parameters
$\xi_2,\ldots,\xi_N$ describe degeneracy for that energy. Hence, for large $N$,
one expects a rather large degeneracy.

Anyway, as is made clear in the quote above, the restrictions on 
$h_{\alpha\beta}$ studied in \cite{BSP} are only relevant when $H$ is {\em not} highly degenerate. 
In other words, the whole BSP argument for violation of either locality or energy-momentum conservation
rests on the assumption that the Hamiltonian is not highly degenerate.
But this, we claim, is a false assumption, so their argument is flawed. 
Indeed, when $H$ is highly degenerate, then there is a large set 
of mutually independent operators which commute with $H$. In this case, the analysis in \cite{BSP}
does not exclude the possibility that all the operators $Q^{\alpha}$ in (\ref{eq1}) belong to this large set, 
in which case the evolution (\ref{eq1}) conserves energy without any restrictions 
on $h_{\alpha\beta}$. We discuss such a conservation of energy and momentum in more detail
in Sec.~\ref{SEC-open}, by studying a more convenient Lindblad form of the evolution equation (\ref{eq1}). 

\section{Physical Hamiltonian is highly degenerate}
\label{SEC-degen}

Why does the Hamiltonian must be degenerate? In essence, because no changes would ever happen
in a universe with a conserved but non-degenerate Hamiltonian. If there was only one state
with given initial energy, then energy conservation would forbid any transition to a new state
different from the initial one. Nothing would ever happen in such a universe, in a clear 
contradiction with observations. 

But how much degenerate a Hamiltonian can be? Can the number of different states with the same energy
be relatively small? In classical mechanics with $N$ degrees of freedom, the total phase space
is $2N$-dimensional, while the phase subspace of a fixed energy $E$ has dimension $2N-1$.
(Indeed, one of the authors of the BSP argument considers
this fact to be a part of the theoretical minimum one needs to know to start doing physics \cite{susskind}.)
For a realistic system with large $N$,
a $(2N-1)$-dimensional space of states with the same energy corresponds to a rather large 
classical degeneracy of the Hamiltonian.

How about quantum systems? When the evolution is unitary, 
a convenient way to describe changes in quantum physics
is to study the $S$-matrix with matrix elements $\langle f|S|i\rangle$.
As explained in any textbook on quantum field theory 
(including the ones \cite{peskin,banks} by the other two 
authors of the BSP argument),
the $S$-matrix conserves energy, i.e. states $|i\rangle$ and $|f\rangle$
are degenerate eigenstates of the Hamiltonian. Indeed, the number of states
with the same energy is very large. For instance, states such as
\begin{equation}\label{degener}
|\omega\rangle, \;\;\; \left|\frac{\omega}{2},\frac{\omega}{2}\right\rangle, \;\;\;
\left|\frac{\omega}{4},\frac{3\omega}{4}\right\rangle, \;\;\; 
\left|\frac{\omega}{3},\frac{\omega}{3},\frac{\omega}{3}\right\rangle, \;\;\; \cdots
\end{equation}
all have the same energy $E=\omega$. (Here $|\omega\rangle$ is a one-particle state
with energy $\omega$, the state $|\frac{\omega}{2},\frac{\omega}{2}\rangle$ is a two-particle state
each of which has energy $\frac{\omega}{2}$, etc.) In a calculation of a 
total cross section, the integration over the quantum phase space of all states allowed by conservation laws
may sometimes even lead to an IR divergence.

Of course, the common energy $E=\omega$ of the states in (\ref{degener}) is the energy of the free
Hamiltonian. Can interactions reduce the degeneracy?
In general, if $|1\rangle$ and $|2\rangle$ are two states with the same free energy $E_0$,
the interaction may cause a split in the energy spectrum, so that $|1\rangle$ and $|2\rangle$ get 
different energies
\begin{equation}
 E_1=E_0+\delta E_1 , \;\;\; E_2=E_0+\delta E_2 ,
\end{equation}
respectively.
Thus, if the measured system is in the superposition $|1\rangle +|2\rangle$ before the interaction,
during the interaction the measured system may end up in one of the states $|1\rangle$ or $|2\rangle$.
However, such a transition must conserve energy of the {\em whole} system, which includes both the measured system
and the source of interaction.
If, for instance, during the interaction the measured system ends up in the state $|1\rangle$, then
its change of energy by $\delta E_1$ must be compensated by a change $-\delta E_1$ in the energy of the source,
so that the total energy does not change. An analogous statement, of course, is also true when the 
measured system ends up in the 
state $|2\rangle$. Hence, even though the subsystem states $|1\rangle$ and $|2\rangle$ may have different energies
due to interaction, the corresponding two states
for the whole system must have the same energy. 
This demonstrates that the interaction does not reduce degeneracy of the total Hamiltonian.

\section{Energy-momentum conservation in open quantum systems}
\label{SEC-open}

In general, energy and momentum of an open system do not need to be conserved; a part of 
energy and momentum can be transferred to the environment. Nevertheless, it does not mean that energy and momentum
can {\em not} be conserved in an open system. In this section we explain in detail how a large degeneracy
of the Hamiltonian implies a possibility of energy-momentum conservation, without violation of locality.

For simplicity we work in the Markov approximation, in which case, as derived in Appendix \ref{app-lindblad},
the density matrix $\rho(t)$ of the open system satisfies the Lindblad equation \cite{lindblad}
\begin{equation}\label{eq-lindblad-dot}
 \dot{\rho}=-i[H,\rho]+\sum_{k\neq 0} \left[ L_k \rho L^{\dagger}_k 
-\frac{1}{2} L^{\dagger}_k L_k \rho -\frac{1}{2} \rho L^{\dagger}_k L_k \right] .
\end{equation}   
The Lindblad equation
is known to be the most general Markovian and time-homogeneous differential equation for $\rho(t)$ \cite{lindblad}.
Indeed, the Lindblad form (\ref{eq-lindblad-dot}) can be obtained
from (\ref{eq1}) by diagonalization of $h_{\alpha\beta}$ \cite{gielen}.
The Lindblad equation is discussed 
in many monographs on quantum mechanics \cite{breuer,decoh1,decoh2,audretsch,nielsen-chuang,laloe}
and even in some general quantum-mechanics textbooks \cite{lebellac,auletta,schumacher}. 

The operators $H$, $L_k$ and $\rho$ appearing in (\ref{eq-lindblad-dot}) act in a Hilbert space ${\cal H}$.
This means that the objects $H$, $L_k$ and $\rho$ live in the product space 
${\cal H}^2\equiv{\cal H}\otimes{\cal H}^*$, where ${\cal H}^*$ is the space dual to ${\cal H}$. 
The Hilbert space ${\cal H}$ can be decomposed as
\begin{equation}
 {\cal H}=\bigoplus_{E}{\cal H}_E ,
\end{equation}
where ${\cal H}_E$ is the Hilbert state of all eigenstates of the Hamiltonian $H$ with the eigenvalue $E$.
In other words, all states in ${\cal H}_E$ have the same energy $E$, and the dimension of ${\cal H}_E$
measures the degeneracy of the Hamiltonian in the sector of energy $E$.
The Hamiltonian can then be written in the spectral form
\begin{equation}\label{Ham}
 H=\sum_E \sum_{\xi} E |E,\xi\rangle \langle E,\xi| ,
\end{equation}
where $|E,\xi\rangle$ are eigenstates of $H$, while $\xi$ is a collective label (compare with (\ref{eq2}))
which labels different states with the same energy $E$. 

Now assume that $\rho(0)\in {\cal H}^2_{E_0}$, i.e. that initially the system has some definite energy $E_0$.
Then (\ref{Ham}) implies $[\rho(0),H]\in {\cal H}^2_{E_0}$, implying that the first term in (\ref{eq-lindblad-dot})
conserves energy. Possible violation of energy-conservation can only arise from the second term in 
(\ref{eq-lindblad-dot}).
However, if each $L_k$ lives in some subspace ${\cal H}^2_E$, or in other words, if for each $L_k$ there is energy $E$
such that
\begin{equation}\label{L-conserved}
 L_k\in {\cal H}^2_E ,
\end{equation}
then also $L^{\dagger}_k\in {\cal H}^2_E$ and we have
\begin{eqnarray}
 L_k \rho(0) L^{\dagger}_k \in{\cal H}^2_{E_0} & {\rm for} \;\; E=E_0 , &
\nonumber \\
 L_k \rho(0) L^{\dagger}_k =0 & {\rm for} \;\; E \neq E_0 , 
\end{eqnarray}
and similarly for other sub-terms in the second term in (\ref{eq-lindblad-dot}).
This shows that the energy is conserved by (\ref{eq-lindblad-dot}) when $L_k$ satisfy
(\ref{L-conserved}). But the constraint (\ref{L-conserved}) allows nontrivial
$L_k$ only when the dimension of ${\cal H}_E$ is larger than one, i.e. only when 
the $E$-sector of the Hamiltonian is degenerate. Hence, {\em when the Hamiltonian is degenerate,
then $L_k$ can be such that (\ref{eq-lindblad-dot}) describes a non-unitary evolution
which conserves energy}.

What about conservation of the 3-momentum ${\bf P}$? We assume that the momentum is conserved
by the purely Hamiltonian evolution, i.e. that
\begin{equation}
 [H,{\bf P}]=0.
\end{equation}
But then ${\bf P}$ has the same eigenstates as $H$ and can be written in the spectral form 
similar to (\ref{Ham})
\begin{equation}\label{P}
{\bf P} =\sum_E \sum_{\xi} {\bf p}(E,\xi) |E,\xi\rangle \langle E,\xi| ,
\end{equation}
where ${\bf p}(E,\xi)$ are eigenstates of the operator ${\bf P}$, i.e. 
${\bf P}|E,\xi\rangle={\bf p}(E,\xi)|E,\xi\rangle$. Thus, denoting by ${\cal H}_{{\bf p}}$
the Hilbert space of all states with the same total momentum ${\bf p}$, we see that $L_k$ can be 
such that they satisfy both (\ref{L-conserved}) and  
\begin{equation}\label{p-conserved}
 L_k\in {\cal H}^2_{{\bf p}} .
\end{equation}
In this case (\ref{eq-lindblad-dot}) conserves both energy and momentum. Hence we see 
that {\em a large degeneracy of the Hamiltonian may lead to conservation of both energy and momentum}.

In \cite{BSP}, where it was assumed that degeneracy is not present, 
it was shown that conservation of energy-momentum is incompatible with locality.
Let us see how the presence of degeneracy avoids that conclusion, i.e.
how both locality and energy-momentum conservation can be present in the presence 
of degeneracy. For that purpose, let us assume that the first sub-term in the
second term in (\ref{eq-lindblad-dot}) has a local form
\begin{equation}\label{local}
 \sum_{k} L_k \rho L^{\dagger}_k = \int d^3x\, {\cal L}({\bf x}) \rho {\cal L}^{\dagger}({\bf x}) ,
\end{equation}
and similarly for the other sub-terms. By introducing the Fourier transforms
\begin{equation}\label{fourier}
{\cal L}({\bf x})=\int \frac{d^3p}{(2\pi)^3} \, e^{-i{\bf p}{\bf x}} L({\bf p}) , \;\;\;
{\cal L}^{\dagger}({\bf x})=\int \frac{d^3p'}{(2\pi)^3} \, e^{i{\bf p'}{\bf x}} L^{\dagger}({\bf p'}) ,
\end{equation}
inserting them in (\ref{local}), and using the identity
\begin{equation}
 \int d^3x\,e^{-i({\bf p}-{\bf p'}){\bf x}} = (2\pi)^3 \delta^3({\bf p}-{\bf p'}) ,
\end{equation}
one obtains
\begin{equation}\label{local-p}
 \sum_{k} L_k \rho L^{\dagger}_k = \int\frac{d^3p}{(2\pi)^3} \, L({\bf p}) \rho L^{\dagger}({\bf p}) .
\end{equation}
How do operators $L({\bf p})$ and $L^{\dagger}({\bf p})$ act on $\rho$? In the absence of degeneracy,
one might argue that they change the total momentum of $\rho$ by ${\bf p}$, which violates 
the conservation of momentum. Indeed, this is essentially
what BSP argued in \cite{BSP} after their Eq. (21) (with an unimportant
difference that they worked with operators $Q^{\alpha}({\bf p})$ instead of our Lindblad operators 
$L({\bf p})$). However, in the presence of degeneracy such an argument is unjustified.
When ${\bf P}$ is degenerated then $L({\bf p})$ can satisfy (\ref{p-conserved}), in which case
the momentum is conserved. Indeed, such operators $L({\bf p})$ can easily be constructed
from any two different states $|{\bf p},\xi\rangle$ and $|{\bf p},\xi'\rangle$
with the same momentum ${\bf p}$, by taking   
\begin{equation}\label{simple-oper}
 L({\bf p}) \propto |{\bf p},\xi\rangle\langle{\bf p},\xi'| . 
\end{equation}
Likewise, when $H$ is degenerated 
then $L({\bf p})$ can satisfy (\ref{L-conserved}), in which case the energy is conserved.
In other words, {\em the degeneracy implies that locality is compatible
with energy-momentum conservation}.

At this point it is also instructive to mention a related error made by Srednicki in \cite{srednicki}.
He claimed that in quantum field theory there are very few operators available which commute with the Hamiltonian, 
and that all of them are global (that is, integrals over all space of a local density).
As the very few examples he had in mind, he mentioned Hamiltonian $H$, momentum ${\bf P}$, and a conserved charge.
But contrary to his claim, we see that there are {\em many} different operators of the form (\ref{simple-oper}) which 
commute with the Hamiltonian (\ref{Ham}). They can be used to construct various local operators 
of the form (\ref{fourier}) which also commute with the Hamiltonian. Finally, they can be 
integrated over space to construct various global operators which also commute with the Hamiltonian.
The high degeneracy of $H$ and ${\bf P}$ opens many possibilities
for construction of various local or global operators which commute with $H$ and ${\bf P}$.   
Of course, unlike the few operators such as $H$, ${\bf P}$ and the conserved charge,
most of these operators cannot be easily expressed in terms of the canonical field operators
and their conjugate momenta. Nevertheless, all these operators share the same physical 
Hilbert space ${\cal H}$ on which they act. 

\section{Open systems with gravity}  
\label{SEC-open-grav}

So far we have shown that energy-momentum in an open system {\em can} be conserved, but we have not claimed
that it {\em must} be conserved. But when gravity is also taken into account, we can strengthen our claims. 
For a certain wide class of open systems involving gravity, 
in this section we argue that energy-momentum of the open system must necessarily be conserved.

Let us start with energy-momentum conservation in classical general relativity. 
Let the metric $g_{\mu\nu}(x)$ approach the Minkowski metric $\eta_{\mu\nu}$ at infinity. By writing
$g_{\mu\nu}(x)=\eta_{\mu\nu}+h_{\mu\nu}(x)$, the Einstein equation can be written 
in an exact but non-covariant form \cite{weinberg}
\begin{equation}\label{einstein}
 R^{(1)}_{\mu\nu}(x)-\frac{1}{2} \eta_{\mu\nu} R^{(1)}(x)=-8\pi G_{\rm N} \tau_{\mu\nu}(x) .
\end{equation}
Here $\tau_{\mu\nu}$ is a pseudo-tensor (i.e.~a tensor under Lorentz transformations,
but not under general coordinate transformations) which satisfies
\begin{equation}\label{local-cons}
 \partial_{\nu}\tau^{\mu\nu}(x)=0.
\end{equation}
All the indices are raised and lowered by $\eta^{\mu\nu}$ and $\eta_{\mu\nu}$,
$R^{(1)}_{\mu\nu}$ is the part of the Ricci tensor linear in $h_{\mu\nu}$, and
\begin{equation}
 \tau_{\mu\nu}(x)=T^{\rm matter}_{\mu\nu}(x)+T^{\rm grav}_{\mu\nu}(x) ,
\end{equation}
where $T^{\rm matter}_{\mu\nu}$ is the matter energy-momentum tensor and $T^{\rm grav}_{\mu\nu}$ 
is the gravitational energy-momentum pseudo-tensor \cite{weinberg}. 
Introducing the global pseudo-vectors
\begin{equation}\label{global-P}
P_{\rm matter}^{\mu}=\int d^3x\, T_{\rm matter}^{\mu 0}(x) , \;\;\;
P_{\rm grav}^{\mu}=\int d^3x\, T_{\rm grav}^{\mu 0}(x) ,
\end{equation}
from (\ref{local-cons}) we see that the total energy-momentum pseudo-vector
\begin{equation}
 P^{\mu}=P_{\rm matter}^{\mu}+P_{\rm grav}^{\mu}
\end{equation}
is conserved. In addition, the
Einstein equation (\ref{einstein}) shows that the total energy-momentum can be interpreted as the source of gravity.

Now consider a compact simply connected 3-dimensional region ${\cal R}$ in space. The whole space consists of the 
interior of ${\cal R}$ and the exterior of ${\cal R}$. Hence the total energy-momentum can be decomposed as
\begin{equation}\label{total-P}
 P^{\mu}=P_{\rm in}^{\mu}+P_{\rm ext}^{\mu} ,
\end{equation}
where $P_{\rm in}^{\mu}$ and $P_{\rm ext}^{\mu}$ are the internal and external energy-momentum,
defined by the corresponding internal and external regions of integration in (\ref{global-P}). 

Now let the open system be defined as the collection of all degrees of freedom in the exterior of ${\cal R}$.
The exterior can exchange energy-momentum with the interior, so $P_{\rm ext}^{\mu}$ 
does not need to be conserved. At first, one might interpret this as non-conservation of energy-momentum
in the open system. However, the interior energy-momentum $P_{\rm in}^{\mu}$ is not really hidden
to the exterior observer. The interior energy-momentum is the {\em source} of gravitational field,
and this gravitational field produced by the interior energy-momentum can be measured even at the 
exterior. Hence, by measuring the external gravitational field, the external observer 
can determine $P_{\rm in}^{\mu}$. (This is similar to classical electrodynamics, where one can determine 
the interior charge by measuring the exterior electromagnetic field.) 
In this way the exterior observer can know not only $P_{\rm ext}^{\mu}$, but also $P_{\rm in}^{\mu}$.
Consequently, the total energy-momentum (\ref{total-P}) is known to the external observer, so 
effectively such an open system can be said to conserve the total energy-momentum. More precisely,
the open system conserves {\em information} about the total energy-momentum, but operationally 
this can be thought of as being the same as conserving the total energy-momentum itself.

Such an operational way of thinking in terms of information is particularly natural
in the context of {\em quantum} information \cite{nielsen-chuang}. 
Indeed, the density matrix $\rho$ of the open quantum 
system describes the {\em information} available in the system. 
Hence, it seems reasonable to expect that in quantum gravity the external density matrix $\rho$
conserves energy-momentum, at least at distances much larger than the Planck length, where the 
Einstein equation (\ref{einstein}) is expected to be a good approximation.

Let us express this expectation in a more quantitative form. Assume that initially the external state
is a pure state $|\psi(0)\rangle_{\rm ext}$ with a definite total energy $E$. At a later time $t$ the external system 
can become entangled with the internal one, so the total state becomes a superposition of the form
\begin{equation}\label{superpos-grav}
 |\Psi(t)\rangle = \sum_{\chi} \sum_{\xi} C_{\chi\xi}(t) \,
|\chi\rangle_{\rm in} |E,\xi\rangle_{\rm ext} ,
\end{equation}
where $C_{\chi\xi}(t)$ are some suitably normalized coefficients.
Due to the conservation of total energy in the exterior, all the external states $|E,\xi\rangle_{\rm ext}$
have the same total energy $E$ equal to the initial exterior energy. 
(For simplicity, the conservation of 3-momentum is not considered explicitly,
but it is not difficult to include it as well.) 
These external states can be further decomposed into the gravitational and matter parts as
\begin{equation}
 |E,\xi\rangle_{\rm ext}=|E_{\rm grav},\xi_{\rm grav}\rangle  |E_{\rm matter},\xi_{\rm matter}\rangle ,
\end{equation}
where $E_{\rm grav}=E-E_{\rm matter}$, so (\ref{superpos-grav}) can 
be written in a more detailed form
\begin{eqnarray}\label{superpos-grav2}
 |\Psi(t)\rangle & = & \displaystyle\sum_{\chi} \sum_{E_{\rm matter}} \sum_{\xi_{\rm grav}} \sum_{\xi_{\rm matter}} 
C_{\chi E_{\rm matter}\xi_{\rm grav}\xi_{\rm matter}}(t) \, |\chi\rangle_{\rm in} 
\nonumber \\
 & &
\otimes |E-E_{\rm matter},\xi_{\rm grav}\rangle  |E_{\rm matter},\xi_{\rm matter}\rangle .
\end{eqnarray} 
The external state $\rho(t)$ is obtained by tracing the pure state $|\Psi(t)\rangle\langle\Psi(t)|$
over the internal degrees of freedom, so (\ref{superpos-grav}) leads to
\begin{equation}\label{mixed-grav}
 \rho(t)=\sum_{\xi} w_{\xi}(t)\, |E,\xi\rangle_{\rm ext} \; _{\rm ext}\langle E,\xi| ,
\end{equation}
where $w_{\xi}(t)= \sum_{\chi}|C_{\chi\xi}(t)|^2$. The state (\ref{mixed-grav}) is a mixed state
with a definite energy $E$. Recalling that the initial external state was the pure state 
$\rho(0)=|\psi(0)\rangle_{\rm ext}\; _{\rm ext}\langle\psi(0)|$ with the same energy $E$, we see that the evolution
(\ref{mixed-grav}) is a {\em non-unitary evolution which conserves energy}.

\section{Application to Hawking radiation}
\label{SEC-hawk}

Now let us apply the general results of the preceding sections to the case of Hawking radiation \cite{hawk1}.
The interior region ${\cal R}$ is the interior of a black hole, which is a special case 
of the situation studied in Sec.~\ref{SEC-open-grav}. For simplicity, we consider only 
spherically symmetric modes of radiation, so the conservation of 3-momentum is trivial.
Hence only the energy conservation is considered explicitly.

\subsection{Energy conservation by Hawking radiation}

Let the initial exterior state be a pure state
\begin{equation}
|\psi(0)\rangle = |M\rangle |0\rangle .
\end{equation}
Here $|M\rangle$ is the external state of gravity the source of which is a black hole with 
initial mass $M$, while $|0\rangle$ is the matter vacuum. After some time $\Delta t$, 
the total state takes the form \cite{nik-epjc}
\begin{equation}\label{superpos-bh}
 |\Psi(\Delta t)\rangle  =  \sum_{E_{\rm rad}} \sum_{\xi_{\rm rad}}  
C_{E_{\rm rad}\xi_{\rm rad}}(\Delta t) \, |-E_{\rm rad},\xi_{\rm rad}\rangle_{\rm in} 
|M-E_{\rm rad}\rangle  |E_{\rm rad},\xi_{\rm rad}\rangle ,
\end{equation} 
which is a special case of (\ref{superpos-grav2}). Here $|M-E_{\rm rad}\rangle$ is the 
external gravitational state corresponding to the black-hole mass $M'=M-E_{\rm rad}$, while 
$E_{\rm rad}$ is the energy of Hawking radiation in the state $|E_{\rm rad},\xi_{\rm rad}\rangle$.
The gravitational state $|M-E_{\rm rad}\rangle$ does not have any degeneracy parameter $\xi_{\rm grav}$,
which is a consequence of the no-hair theorem. 
(This theorem \cite{hawking-ellis,frolov-novikov}
states that the gravitational field external to the stationary 
black hole reveals no other information about the black hole except its mass, charge and angular momentum,
and for simplicity we assume that charge and angular momentum are zero.)
The external states of Hawking radiation $|E_{\rm rad},\xi_{\rm rad}\rangle$ and the internal states 
$|-E_{\rm rad},\xi_{\rm rad}\rangle_{\rm in}$ are parameterized by the same parameters 
$E_{\rm rad}$ and $\xi_{\rm rad}$, which corresponds to the perfect correlation of
external Hawking radiation with internal Hawking quanta. 
(Of course, the internal Hawking quanta are not the only quantum degrees in the interior. 
However, the other internal degrees will not play explicit role in our calculations, so we suppress them.)
The Hawking-radiation states $|E_{\rm rad},\xi_{\rm rad}\rangle$
with the same $E_{\rm rad}$ but different $\xi_{\rm rad}$ include different states of the form (\ref{degener}).

The Hawking radiation has a thermal spectrum, which means that
\begin{equation}\label{norm1}
 |C_{E_{\rm rad}\xi_{\rm rad}}(\Delta t)|^2= \frac {e^{-\beta E_{\rm rad}}} {Z(\Delta t)} .
\end{equation}
Here $Z(\Delta t)$ is a normalization constant to be discussed later and $\beta^{-1}=T$ 
is the Hawking temperature \cite{hawk1}
\begin{equation}\label{T}
 T=\frac{1}{8\pi M} ,
\end{equation}
in the natural units $\hbar=c=G_{\rm N}=k_{\rm B}=1$.
The external state associated with (\ref{superpos-bh}) is the mixed thermal state
\begin{equation}\label{mixed-bh}
 \rho(\Delta t)=\sum_{E_{\rm rad}} \sum_{\xi_{\rm rad}}  |C_{E_{\rm rad}\xi_{\rm rad}}(\Delta t)|^2 \,
|M,E_{\rm rad},\xi_{\rm rad}\rangle \langle M,E_{\rm rad},\xi_{\rm rad}| ,
\end{equation}
where
\begin{equation}\label{single-notation}
|M,E_{\rm rad},\xi_{\rm rad}\rangle \equiv |M-E_{\rm rad}\rangle  |E_{\rm rad},\xi_{\rm rad}\rangle
\end{equation}
are states all having the same total external energy $M$.  

\subsection{Violation of unitarity}

The evolution (\ref{superpos-bh}) is unitary, which is related to the fact that it is obtained 
with the aid of a Bogoliubov transformation \cite{sidorov}. 
However, the creation of internal Hawking quanta $|-E_{\rm rad},\xi_{\rm rad}\rangle_{\rm in}$
increases the internal entanglement entropy of the black hole. In this way 
the black hole entropy can easily become larger than the Bekenstein-Hawking entropy
$S=A/4$, where $A=4\pi R^2$ is the area of the black-hole horizon at the radius $R=2M$.
This is a problem, not only because there are many reasons to believe that $S=A/4$ should be the correct
black-hole entropy, but even worse, because in this way the black-hole evaporation could eventually
create a very small and light black hole with an arbitrarily large amount of entropy. 
This is one of the ways to formulate the infamous black-hole information paradox.

This form of the paradox can be removed by assuming that the internal Hawking quanta 
are destroyed by a non-unitary process near the black-hole singularity. But then the external mixed state
(\ref{mixed-bh}) becomes a state of a {\em closed} system. That means that a closed system
evolves non-unitarily from a pure to a mixed state, which
the standard quantum theory does not allow.
This represents another form of the black-hole information paradox. 

Of course, it is conceivable that unitarity of standard quantum theory might break down
at black-hole singularities.
But, as said in the Introduction,   
one of the most frequent arguments against non-unitary evolution for the closed system is 
the BSP argument \cite{BSP} that such evolution would be incompatible with either energy-momentum conservation
or locality.
Yet, in this paper we have shown that such an argument is incorrect. Indeed,
the non-unitary evolution (\ref{mixed-bh}), now interpreted as the evolution for a 
closed system, conserves energy-momentum. On the other hand, the creation of Hawking
particles at the horizon is a local process \cite{nik-part1,nik-part2,nik-part3}.
If, as we suggested above, the true violation of unitarity happens
near the black-hole singularity where the internal Hawking quanta are destroyed, 
then the violation of unitarity does not seem to violate locality, or at least not
at distances much larger than the Planck length. 

\subsection{Lindblad operators for Hawking radiation}

The general argument in Sec.~\ref{SEC-open} 
for compatibility between non-unitarity and energy-momentum conservation involved the 
Lindblad equation. To see the connection between (\ref{mixed-bh}) and the
Lindblad equation, let us now calculate the Lindblad operators corresponding
to (\ref{mixed-bh}). 

For our case, Eq.~(\ref{kraus-action}) in Appendix \ref{app-lindblad} can be written as
\begin{equation}
_{\rm in}\langle -E_{\rm rad},\xi_{\rm rad}|\Psi(\Delta t)\rangle  
= A_{E_{\rm rad}\xi_{\rm rad}}(\Delta t) |M\rangle |0\rangle .
\end{equation}
Using (\ref{superpos-bh}), this gives
\begin{equation}
 A_{E_{\rm rad}\xi_{\rm rad}}(\Delta t) |M\rangle |0\rangle 
= C_{E_{\rm rad}\xi_{\rm rad}}(\Delta t) |M-E_{\rm rad}\rangle  |E_{\rm rad},\xi_{\rm rad}\rangle ,
\end{equation}
which in the notation of (\ref{single-notation}) can be written as
\begin{equation}
 A_{E_{\rm rad}\xi_{\rm rad}}(\Delta t) |M,0,0\rangle 
= C_{E_{\rm rad}\xi_{\rm rad}}(\Delta t) |M,E_{\rm rad},\xi_{\rm rad}\rangle .
\end{equation}
Hence the Kraus operators $A_{E_{\rm rad}\xi_{\rm rad}}(\Delta t)$ are
\begin{equation}\label{kraus-hawk}
 A_{E_{\rm rad}\xi_{\rm rad}}(\Delta t) = C_{E_{\rm rad}\xi_{\rm rad}}(\Delta t)
|M,E_{\rm rad},\xi_{\rm rad}\rangle \langle M,0,0|.
\end{equation}

In (\ref{superpos-bh}), $|C_{E_{\rm rad}\xi_{\rm rad}}(\Delta t)|^2$ is the probability 
that the system will be found in the corresponding state after the time $\Delta t$.
For a relatively short time $\Delta t$ 
(but not shorter than the Zeno time \cite{nik-zeno}),
this probability increases linearly with time. Therefore we can write
\begin{equation}\label{norm2}
 |C_{E_{\rm rad}\xi_{\rm rad}}(\Delta t)|^2= \Delta t |C'_{E_{\rm rad}\xi_{\rm rad}}|^2 ,
\end{equation}
where $|C'_{E_{\rm rad}\xi_{\rm rad}}|^2$ is the transition probability per unit time
which does not depend on $\Delta t$. 
Hence (\ref{norm1}) and (\ref{norm2}) imply 
\begin{equation}\label{norm3}
 |C'_{E_{\rm rad}\xi_{\rm rad}}|^2= \frac {e^{-\beta E_{\rm rad}}} {Z'} ,
\end{equation}
where $Z'=\Delta t \, Z(\Delta t)$ does not depend on $\Delta t$. 

To find the constant
$Z'$, we calculate the total average power of Hawking radiation 
\begin{equation}\label{power}
 \left\langle \frac{dE_{\rm rad}}{dt} \right\rangle = \sum_{E_{\rm rad}} \sum_{\xi_{\rm rad}} 
E_{\rm rad} |C'_{E_{\rm rad}\xi_{\rm rad}}|^2 
= \frac{1}{Z'}  \sum_{E_{\rm rad}} \sum_{\xi_{\rm rad}} E_{\rm rad} \, e^{-\beta E_{\rm rad}}.
\end{equation}
The left-hand side of (\ref{power}) can also be calculated from the Stefan-Boltzmann law of black-body radiation
\begin{equation}
 \left\langle \frac{dE_{\rm rad}}{dt} \right\rangle = \sigma A T^4 ,
\end{equation}
where $\sigma=\pi^2/60$ is the Stefan-Boltzmann constant, $A=4\pi R^2=16\pi M^2$ is the black-hole area,
and $T$ is the Hawking temperature (\ref{T}). This gives
\begin{equation}\label{sigma'}
 \left\langle \frac{dE_{\rm rad}}{dt} \right\rangle = \frac{\sigma'}{M^2} ,
\end{equation} 
where $\sigma'=\sigma/256\pi^3$. Hence (\ref{power}) implies that the normalization constant $Z'$ is 
\begin{equation}
 Z'= \frac{ \sum_{E_{\rm rad}} \sum_{\xi_{\rm rad}} E_{\rm rad} \, e^{-\beta E_{\rm rad}} } 
{\sigma'/M^2}.
\end{equation}

Now the Lindblad operators can be calculated by applying the general equation (\ref{two-eqs}) 
in Appendix \ref{app-lindblad}
to our specific equation (\ref{kraus-hawk}). Using also (\ref{norm2}) and (\ref{norm3}), this finally gives
\begin{equation}\label{lind-hawk}
 L_{E_{\rm rad}\xi_{\rm rad}} = \frac{e^{-\beta E_{\rm rad}/2}}{\sqrt{Z'}} \, 
|M,E_{\rm rad},\xi_{\rm rad}\rangle \langle M,0,0| .
\end{equation}
Note that the phases of the Lindblad operators are irrelevant, because the phases cancel out
in (\ref{eq-lindblad-dot}), in products involving both $L_k$ and $L^{\dagger}_k$.
As the states $|M,E_{\rm rad},\xi_{\rm rad}\rangle$ and $|M,0,0\rangle$ both have the same total
energy $M$ (recall Eq.~(\ref{single-notation})), 
we see that {\em the Lindblad operators (\ref{lind-hawk}) conserve energy}. 
(The conservation of 3-momentum is trivial, as explained in the introductory paragraph
of Sec.~\ref{SEC-hawk}.)

We also see that the Lindblad operators (\ref{lind-hawk}) depend on the black-hole mass $M$. 
Therefore, they can be treated as time-independent operators as long as the black-hole mass does not change much.
This is a very good approximation for quite a long time, because (\ref{sigma'}) implies that 
mass of the black hole changes significantly only after a time of the order of $M^3$. Indeed, denoting 
by $\overline{M}(t)$ the average black-hole mass at time $t$, we see that (\ref{sigma'}) and the conservation 
of total average energy imply
\begin{equation}\label{bh-mass}
 \frac{d\overline{M}(t)}{dt} = - \frac{\sigma'}{\overline{M}^2(t)} .
\end{equation}
The solution of this differential equation is
\begin{equation}\label{bh-mass2}
 \overline{M}(t)=\sqrt[3]{M_0^3-3\sigma't} ,
\end{equation}
where now the initial black-hole mass is denoted by $M_0$. 

In this way the Lindblad equation (\ref{eq-lindblad-dot}) 
can be improved by replacing the time-independent Lindblad operators $L_k$ with the time-dependent ones $L_k(t)$.
The time-dependent Lindblad operators are
given by (\ref{lind-hawk}) in which $M$ is replaced by $\overline{M}(t)$. Of course, this is still an 
approximation, but now a better one. In particular, the resulting master equation is still Markovian,
but not longer time-homogeneous. In addition, the resulting time-dependent Lindblad operators still
conserve energy and momentum.
 
\section{Discussion and outlook}
\label{SEC-disc} 

In this paper we have argued that one of the main alleged problems with non-unitary Hawking radiation,
the problem of violation of either locality or conservation of energy-momentum, does not really exist. 
Owing to a high degeneracy of the Hamiltonian and momentum operators, 
non-unitary Hawking radiation is fully compatible with both locality and energy-momentum conservation. 
This eliminates one important obstacle for the possibility that Hawking radiation might really violate unitarity.  

This, however, does {\em not} imply that unitarity violation by Hawking radiation is without problems.
There are also other good reasons (not eliminated by results of the present paper)
to think that evolution should be unitary.

The first reason is the fact that known non-gravitational quantum theories are unitary, suggesting that 
quantum gravity should not be different. Nevertheless, gravity {\em is} different by being a theory
of space and time. In particular, unitary evolution requires a globally hyperbolic 
spacetime, while general relativity contains solutions which are not globally hyperbolic.
This suggests the possibility that, in the presence of gravity, quantum theory should be
generalized to a theory in which the notion of unitarity does not need to refer to a 
unitary time evolution \cite{hartle,nik1,nik2,nik3}. In this way the non-unitary time evolution 
of Hawking radiation may not be in contradiction with a generalized concept of unitarity.

The second reason, especially popular in the string-theory community, is the AdS/CFT correspondence
\cite{maldacena}, or more generally the gauge/gravity correspondence \cite{horowitz}.
According to this correspondence, the quantum theory of gravity is mathematically equivalent 
to a quantum field theory on the boundary. Since the latter is known to be unitary,
it follows that the former must also be. This, indeed, is a very strong argument for
unitarity of Hawking radiation, but is not without possible loopholes.
First, many aspects of this correspondence depend on the assumption of string theory,
while we still do not have an experimental proof that string theory is the correct theory of Nature.
Second, many aspects of the correspondence depend on the assumption of supersymmetry 
and AdS background, while the Universe we observe does not look supersymmetric and we do not 
seem to live on an AdS background. In the absence of supersymmetry and AdS background,
the present evidence for the correspondence is still incomplete.
Third, even if these problems are merely technical and the correspondence is exact 
even without supersymmetry and AdS background, there is an intriguing suggestion that
the field theory part of the correspondence may in fact be a {\em non}-unitary gauge theory \cite{vafa}.

To conclude, the question of unitarity of Hawking radiation seems to be an open
question. In our opinion, non-unitary Hawking radiation is still a viable possibility 
worthwhile of further investigation.

\section*{Acknowledgements}

This work was supported by the Ministry of Science of the Republic of Croatia.

\appendix

\section{Lindblad equation}
\label{app-lindblad}

Since many intended readers of this paper are probably not familiar with the Lindblad equation, 
for the sake of completeness we present here some basics, by giving simple
derivations and explanations of the relevant equations used in the rest of the paper. 
Our derivations are to a large extent based
on \cite{schumacher}, were some further illuminating details can be found. 

Let $|\Psi(t)\rangle$ be the full time-dependent state of a closed system, and let the system be split
into two subsystems, one of which is observed and the other unobserved.
Let $|k\rangle$ be a basis for the unobserved subsystem. Then the density matrix of the observed subsystem is
\begin{equation}\label{app1}
\rho(t) =  {\rm Tr}_{\rm unobs} |\Psi(t)\rangle \langle \Psi(t)| 
= \sum_k \langle k|\Psi(t)\rangle \langle \Psi(t)|k\rangle .
\end{equation}
This can be written as 
\begin{equation}\label{kraus}
\rho(t) = \sum_k A_k(t) \rho(0) A^{\dagger}_k(t), 
\end{equation}
where $A_k(t)$ are the so-called Kraus operators which act only in the Hilbert space 
of the observed subsystem. In particular, if the 
observed subsystem is initially in the pure state $\rho(0)=|\psi(0)\rangle\langle\psi(0)|$,
from (\ref{app1}) and (\ref{kraus}) one can see that
the action of Kraus operators is given by
\begin{equation}\label{kraus-action}
 \langle k|\Psi(t)\rangle=A_k(t) |\psi(0)\rangle .
\end{equation}
In general, the density matrix must satisfy ${\rm Tr}\rho(t)=1$ for any $\rho(0)$, 
so (\ref{kraus}) leads to the constraint
\begin{equation}\label{Kraus-constraint}
 \sum_k A^{\dagger}_k(t) A_k(t) =1.
\end{equation}

Now consider $\rho(\Delta t)$ at a small time $\Delta t$, so that 
\begin{equation}
 \rho(\Delta t)=\rho(0)+\Delta\rho.
\end{equation}
Assuming that
one of the Kraus operators (call it $A_0$) is close to the unit operator, we can write
\begin{eqnarray}\label{two-eqs}
& A_0(\Delta t)=1+\Delta t (L_0-iH), &
\nonumber \\ 
& A_k(\Delta t)= \sqrt{\Delta t} L_k \;\;{\rm for}\;\; k\neq 0,
\end{eqnarray}
where $L_0$ and $H$ are some hermitian operators.
In this way, in the lowest order in $\Delta t$, (\ref{kraus}) leads to the equation
\begin{equation}\label{lindblad0}
 \rho(0)+\Delta\rho = \rho(0)+\Delta t 
\left( \{L_0,\rho(0)\}-i[H,\rho(0)]+\sum_{k\neq 0} L_k \rho(0) L^{\dagger}_k \right).   
\end{equation}

Strictly speaking, this equation is correct only initially, i.e. for $\rho(0)$.
Nevertheless, if the density matrix $\rho(t)$ quickly ``forgets'' its history,
or equivalently, if the time evolution of the observed subsystem 
is sufficiently slow so that there is enough time to ``forget'' 
the history, then one can use the Markov approximation, i.e. one can 
consider (\ref{lindblad0}) to be approximately valid for $\rho(t)$ at {\em any} $t$.

Or let us be slightly more precise \cite{schumacher}. When the unobserved degrees interact with the observed ones,
the interaction usually creates entanglement between them, which makes $\rho$ mixed. 
However if, after the creation of entanglement, the unobserved degrees are removed 
from the vicinity of the observed ones, then the unobserved degrees
have no further influence on the evolution of the observed degrees.
In this way the density matrix $\rho$ at a given time $t$ effectively ``forgets'' that it was mixed 
{\em due to entanglement} with the unobserved degrees, and only ``remembers'' that it is now mixed. In other words, 
all the information relevant for further evolution of $\rho$ is contained in the $\rho(t)$ itself,
not in its environment.
 
In this approximation, and in the limit $\Delta t\rightarrow 0$, 
(\ref{lindblad0}) becomes the differential equation
\begin{equation}\label{pre-lindblad}
 \frac{d\rho(t)}{dt}=\{L_0,\rho(t)\}-i[H,\rho(t)]+\sum_{k\neq 0} L_k \rho(t) L^{\dagger}_k .
\end{equation}
Furthermore, the condition ${\rm Tr}\rho(t)=1$ implies 
\begin{equation}
{\rm Tr}\frac{d\rho(t)}{dt}=0 , 
\end{equation}
so from (\ref{pre-lindblad}) one finds a constraint similar to (\ref{Kraus-constraint})
\begin{equation}
 2L_0+\sum_{k\neq 0} L^{\dagger}_k L_k =1.
\end{equation}
This can be used to eliminate $L_0$ from (\ref{pre-lindblad}), so (\ref{pre-lindblad}) can be written
in the final form
\begin{equation}\label{eq-lindblad}
 \frac{d\rho}{dt}=-i[H,\rho]+\sum_{k\neq 0} \left[ L_k \rho L^{\dagger}_k 
-\frac{1}{2} L^{\dagger}_k L_k \rho -\frac{1}{2} \rho L^{\dagger}_k L_k \right] .
\end{equation}
This is a master equation known as the {\em Lindblad equation} \cite{lindblad}, while
the operators $L_k$ are called Lindblad operators.  
 
We have not yet explained the physical interpretation of the operator $H$. All we know at this moment
is that it is a hermitian operator acting in the Hilbert space of the observed subsystem. 
In general, the evolution (\ref{eq-lindblad}) is not unitary. But in the special case in which all
$L_k=0$ for $k\neq 0$, (\ref{eq-lindblad}) reduces to the familiar unitary evolution 
generated by $H$. This shows that $H$ can be interpreted as the Hamiltonian 
for the observed subsystem. However, $H$ depends also on the interaction with the
unobserved subsystem, meaning that $H$ is a modified Hamiltonian that includes 
the unitary influences of the unobserved subsystem to the observed one. For example,
when a laser beam is reflected at a heavy mirror, it is a unitary process for the beam \cite{schumacher}. 
So if $\rho$ is the state of the laser beam, then
$H$ is an operator acting only in the Hilbert space of the beam, in which case 
$H$ contains both the free Hamiltonian of the beam and a term describing the influence of the mirror.

\end{document}